\documentclass[preprint,eqsecnum,prb]{revtex4}
\usepackage{epsfig,amsmath,bbm,graphicx}

\newcommand{\ee}{\end{equation}}
\newcommand{\be}{\begin{equation}} 
\newcommand{\bea}{\begin{eqnarray}}
\newcommand{\eea}{\end{eqnarray}}
\newcommand{\bml}{\begin{subequations}} 
\newcommand{\eml}{\end{subequations}}

\begin{document}

\title{Numerically exact, time-dependent study of
  correlated electron transport in model molecular junctions}

\author{Haobin Wang}
\affiliation{Beijing Computational Science Research Center,
No. 3 He-Qing Road, Hai-Dian District, Beijing 100084, P.R. China}
\affiliation{Department of Chemistry and Biochemistry, MSC 3C, New Mexico State
University, Las Cruces, NM 88003, USA}

\author{Michael Thoss}
\affiliation{Institute for Theoretical Physics and Interdisciplinary Center for Molecular Materials,
  Friedrich-Alexander-Universit\"at Erlangen-N\"urnberg,
  Staudtstr.\ 7/B2, D-91058, Germany}

\date{\today}

\begin{abstract}
\baselineskip6mm

The multilayer multiconfiguration time-dependent Hartree theory within second 
quantization representation of the Fock space is applied to study correlated
electron transport in models of single-molecule junctions. Extending previous work, 
we consider models which include both electron-electron and electronic-vibrational
interaction. The results show the influence of the interactions on the transient and the
stationary electrical current. The underlying physical mechanisms are
analyzed in conjunction with the nonequilibrium electronic population of the molecular bridge.

\end{abstract}
\maketitle

\section{Introduction}

The process of charge transport in molecular  junctions  has received  much
attention recently.\cite{ree97:252,joa00:541,Nitzan01,nit03:1384,Cuniberti05,Selzer06,Venkataraman06,Chen07,Galperin08b,Cuevas10}
Single molecule junctions, consisting of single
molecules that are chemically bound to metal electrodes, are well-suited systems to study nonequilibrium
transport phenomena at the nanoscale and are also of interest for potential
applications in the field of molecular electronics. Recent
developments in experimental techniques, such as electromigration, mechanically controllable 
break junctions, or scanning tunneling microscopy,\cite{ree97:252,par00:57,cui01:571,par02:722,smi02:906,rei02:176804,zhi02:226801,xu03:1221,qiu04:206102,liu04:11371,elb05:8815,Elbing05,Ogawa07,Schulze08,Pump08,Leon08,Osorio10,Tao10,Martin10} have made it possible to study transport properties of
molecular junctions. The rich experimental observations, e.g.,
Coulomb blockade,\cite{par02:722} Kondo effect,\cite{lia02:725} negative differential 
resistance,\cite{che99:1550,Gaudioso00,Osorio10} switching and hysteresis,\cite{blu05:167,Riel06,Choi06}
have stimulated many theoretical developments for understanding quantum
transport at the molecular scale.

A particular challenge for the theory of charge transport in
molecular junctions is the accurate treatment of correlation effects beyond
the mean-field level. In molecular junctions, there are two types
of correlation effects due to electronic-vibrational and electron-electron interaction.
Considering  vibrational induced correlation effects, a 
variety of theoretical approaches have been developed, including scattering
theory,\cite{Bonca95,Ness01,Cizek04,Cizek05,Toroker07,Benesch08,Zimbovskaya09,Seidemann10} 
nonequilibrium Green's function approaches,\cite{Flensberg03,Mitra04,Galperin06,Ryndyk06,Frederiksen07,Tahir08,Haertle08,Stafford09,Haertle09} 
and master equation
methods.\cite{May02,Mitra04,Lehmann04,Pedersen05,Harbola06,Zazunov06,Siddiqui07,Timm08,May08,May08b,Leijnse09,Esposito09,Volkovich11,Haertle11}
In spite of the physical insight offered by these methods, 
all of them involve significant approximations. For example, NEGF methods and
master equation approaches are usually based on (self-consistent) perturbation theory and/or employ
factorization schemes. Scattering theory approaches to vibrationally coupled electron transport, 
on the other hand, neglect vibrational nonequilibrium effects and are limited to the treatment of a 
small number of vibrational degrees of freedom.
These shortcomings have motivated us to develop a
systematic, numerically exact methodology to study quantum dynamics and quantum transport including 
many-body effects --- the multilayer multiconfiguration time-dependent Hartree (ML-MCTDH) theory 
in second quantization representation (SQR).\cite{wan09:024114} For a generic
model of vibrationally coupled
electron transport, we have demonstrated the importance of treating 
electronic-vibrational coupling accurately. Comparison with approximate methods such as NEGF reveals the
necessity of employing accurate methods such as the ML-MCTDH-SQR, in particular in the strong coupling 
regime.

In this paper, we extend the ML-MCTDH-SQR method to treat electron-electron
interaction. Considering the paradigmatic Anderson impurity model, we show
the applicability of the methodology to obtain an accurate
description. Furthermore, we consider a model which incorporates both
electron-electron and electronic-vibrational interaction. To the best of our
knowledge, the results reported for this model are the first obtained by a
numerically exact method.

It is noted that a variety of other powerful methods have been developed in
the recent years with the same goal, i.e., to facilitate numerically
exact simulations for nonequilibrium transport in model systems. These include
the numerical path  integral approach,\cite{muh08:176403,wei08:195316,Segal10} real-time 
quantum Monte Carlo simulations,\cite{Werner09,Schiro09} the numerical renormalization 
group approach,\cite{and08:066804}, the time-dependent density matrix renormalization 
group approach.\cite{HeidrichMeisner09}, and the hierarchical equations of motion
method \cite{Zheng09,Jiang12}. For a comparison and an comprehensive overview of various different
methods in the case of nonequilibrium transport with
electron-electron interaction see Ref.~\onlinecite{Eckel10}.

The remaining part of the paper is organized as follows. 
Section~\ref{modeltight} outlines the physical model and the observables of interest.  
The ML-MCTDH-SQR theory is described in Section~\ref{mlsqr}.  Section~\ref{results} 
presents numerical results for a variety of parameter regimes as well as an analysis of the 
transport mechanisms. Section~\ref{conclusions} concludes with a summary.

\section{Model and Observables of Interest}\label{modeltight}

To study correlated electron transport in molecular junctions, we consider a
generic model  which includes both electron-electron and
electronic-vibrational
interaction. The model comprises two discrete electronic states (spin up and
down) at the 
molecular bridge, 
two electronic continua describing the left and the right metal leads, respectively, and a distribution 
of harmonic oscillators that models the vibrational modes of the molecular
bridge.
 The Hamiltonian reads
\begin{subequations}\label{Htot}
\begin{equation}
	\hat H = \hat H_{\rm el} + \hat H_{\rm nuc} + \hat H_{\rm el-nuc},
\end{equation}
where $\hat H_{\rm el}$, $\hat H_{\rm nuc}$, and $\hat H_{\rm el-nuc}$ describe the electronic 
degrees of freedom, the nuclear vibrations, and their coupling terms, respectively
\begin{eqnarray}\label{H1tot}
	\hat H_{\rm el} &=& \sum_{\sigma} E_d \hat{n}_{d,\sigma} 
	+ U_d \hat{n}_{d,\uparrow} \hat{n}_{d,\downarrow}
	+ \sum_{k_L,\sigma} E_{k_L} \hat{n}_{k_L,\sigma}
	+ \sum_{k_R,\sigma} E_{k_R} \hat{n}_{k_R,\sigma} \\
  &&	+ \sum_{k_L,\sigma} V_{dk_L} ( \hat{d}^+_\sigma \hat{c}_{k_L,\sigma} + \hat{c}_{k_L,\sigma}^+ \hat{d}_\sigma )
	+ \sum_{k_R,\sigma} V_{dk_R} ( \hat{d}^+_\sigma \hat{c}_{k_R,\sigma} + \hat{c}_{k_R,\sigma}^+ \hat{d}_\sigma ), \nonumber 
\end{eqnarray}
\begin{equation}\label{Hnuc}
	\hat H_{\rm nuc} = \frac{1}{2} \sum_j ( \hat{P}_j^2 + \omega_j^2 \hat{Q}_j^2 ), \\
\end{equation}
\begin{equation}
	\hat H_{\rm el-nuc} =  \sum_{\sigma} \hat{n}_{d,\sigma} \sum_j 2 c_j \hat{Q}_j.
\end{equation}
\end{subequations}
In the above expression $\hat{n}$ denotes the number operator, subscript ``$d$'' refers to the 
bridge state, ``$k_L/k_R$'' the states of the left/right metal leads, and ``$\sigma=\uparrow,\downarrow$''
the two spin states. Operators $\hat{d}^+/ \hat{d}$, $\hat{c}_{k_L}^+/ \hat{c}_{k_L}$, $\hat{c}_{k_R}^+/ \hat{c}_{k_R}$ are the
fermionic creation/annihilation operators for the electronic states on the molecular bridge, the left
and the right leads, respectively. The second term in (\ref{H1tot}) describes
the on-site Coulomb repulsion of the electrons on the molecular bridge 
with electron-electron coupling strength $U_d$. The energies of the electronic
states in the leads, $E_{k_L}$, $E_{k_R}$, as well as the molecule-lead
coupling parameters  $V_{dk_L}$, $V_{dk_R}$ are  assumed to be independent
of the spin polarization and  are defined through the energy-dependent
level width functions
\begin{equation}
	\Gamma_L (E) = 2\pi \sum_{k_L} |V_{dk_L}|^2 \delta(E-E_{k_L}), \hspace{1cm}
	\Gamma_R (E) = 2\pi \sum_{k_R} |V_{dk_R}|^2 \delta(E-E_{k_R}).
\end{equation}

Without electronic-vibrational
interaction the model introduced above  reduces to the well known Anderson
impurity model,\cite{Anderson61} which has been investigated in great detail
both in equilibrium and nonequilibrium.\cite{Hewson93,Eckel10} Neglecting, on
the other hand, electron-electron interaction,  it corresponds to the
standard model of vibrationally coupled electron transport in molecular
junctions, which has also been studied in great detail, mostly based on approximate
methods. Recently a numerically exact treatment of the latter model
(i.e.\ without  electron-electron interaction) became possible
using path integral techniques,\cite{muh08:176403,Albrecht12} as well as the
ML-MCTDH approach.\cite{wan09:024114,Wang11,Albrecht12}
To the best of our knowledge, the full model including electron-electron and
electronic-vibrational interaction has so far not been considered with
numerically exact methods.

In principle, the parameters of the model can be obtained for a specific
molecular junction employing first-principles electronic structure
calculations.\cite{Benesch09} In this paper, which focuses on the methodology
and general transport
properties, however, we will use a generic parameterization.
Employing a tight-binding model, the function $\Gamma (E)$ is given as
\begin{subequations}
\begin{equation}
        \Gamma (E) = \left\{ \begin{array}{ll} \frac{\alpha_e^2}{\beta_e^2} \sqrt{4\beta_e^2-E^2} 
		\hspace{1cm} & |E| \leq 2 |\beta_e| \\
                0 \hspace{1cm} &  |E| > 2 |\beta_e| \end{array}  \right.,
\end{equation}
\begin{equation}
	\Gamma_L (E) = \Gamma (E-\mu_L), \hspace{1cm}  \Gamma_R (E) = \Gamma (E-\mu_R),
\end{equation}
\end{subequations}
where $\beta_e$ and $\alpha_e$ are nearest-neighbor couplings between two lead sites 
and between the lead and the bridge state, respectively.  I.e., the width functions for 
the left and the right leads are obtained by shifting $\Gamma(E)$ relative to the chemical potentials 
of the corresponding leads.  We consider a simple model of two identical leads, in which the chemical 
potentials are given by 
\begin{equation}
	\mu_{L/R} = E_f \pm V/2,
\end{equation}
where $V$ is the source-drain bias voltage and $E_f$ the Fermi energy of the leads. Since only the 
difference $E_d - E_f$ is physically relevant, we set $E_f = 0$.

Similarly, the frequencies $\omega_j$ and electronic-nuclear coupling
constants $c_j$ of the vibrational modes of the molecular junctions are modeled by a spectral density 
function\cite{leg87:1,Weiss93}
\begin{equation}
\label{discrete}
        J(\omega) = \frac{\pi} {2} \sum_{j} \frac{c_{j}^{2}} {\omega_j}
        \delta(\omega - \omega_{j}).
\end{equation}
In this paper, the spectral density is chosen in Ohmic form with an exponential cutoff
\begin{equation}
\label{ohmic}
        J_{\rm O}(\omega)  = \frac{\pi}{2} \alpha \omega e^{-\omega/\omega_c},
\end{equation}
where $\alpha$ is the dimensionless Kondo parameter.  

Both the electronic and the vibrational continua can be discretized by choosing a 
density of states $\rho_e(E)$ and a density of frequencies 
$\rho(\omega)$ such that\cite{tho01:2991,wan01:2979,wan03:1289}
\begin{subequations}
\begin{equation}
	\int_0^{E_k} dE \; \rho_e(E) = k, \hspace{.5in} 
	|V_{dk}|^2 =  \frac{\Gamma(E_k)}{2\pi \rho_e(E_k)}, \hspace{.5in}
	 k = 1,...,N_e,
\end{equation}
\begin{equation}
	\int_0^{\omega_j} d\omega \; \rho(\omega) = j, \hspace{.5in}  
	\frac{c_{j}^{2}} {\omega_j} = \frac{2}{\pi} \frac{J_{\rm O}(\omega_j)}{\rho(\omega_j)},
 	\hspace{.5in} j = 1,...,N_b.
\end{equation}
\end{subequations}
where $N_e$ is the number of electronic states (for a single spin/single lead) and $N_b$ is the number of
bath modes in the simulation.  In this work, we choose a constant $\rho_e(E)$, i.e., an equidistant
discretization of the interval $[-2\beta_e, 2\beta_e]$, to discretize the electronic continuum.  For 
the vibrational bath, $\rho(\omega)$ is chosen as
\begin{equation}
        \rho(\omega) = \frac{N_b+1}{ \omega_c} e^{-\omega/\omega_c}.
\end{equation} 
Within a given time scale the numbers of electronic states and bath 
modes are systematically increased to reach converged results for the quantum dynamics in the 
extended condensed phase system.  In our calculations we employ 80-500 states for each electronic lead, implying 
40-250 electrons per lead, and a bath with 100-400 modes.

The observable of interest for studying transport through molecular junctions is the current for 
a given source-drain bias voltage, given by (in this paper we use atomic units where $\hbar = e = 1$)
\begin{subequations}
\begin{equation}
	I_L(t) = - \frac{d N_L(t)} {dt} = -\frac{1}{{\rm tr}[\hat{\rho}]} {\rm tr}
        \left\{ \hat{\rho} e^{i\hat{H}t} i[\hat{H}, \hat{N}_{L}] e^{-i\hat{H}t} \right\},
\end{equation}
\begin{equation}
	I_R(t) = \frac{d N_R(t)} {dt} = \frac{1}{{\rm tr}[\hat{\rho}]} {\rm tr}
        \left\{ \hat{\rho} e^{i\hat{H}t} i[\hat{H}, \hat{N}_{R}] e^{-i\hat{H}t} \right\}.
\end{equation}
\end{subequations}
Here, $N_{L/R}(t)$ denotes the  time-dependent charge in each lead, defined as 
\begin{equation}
	N_{\zeta}(t) = \frac{1}{{\rm tr}[\hat{\rho}]} {\rm tr}
        [\hat{\rho} e^{i\hat{H}t} \hat{N}_{\zeta} e^{-i\hat{H}t} ], \;\;\; \zeta=L, R,
\end{equation}
and  $\hat{N}_{\zeta} = \sum_{k_\zeta,\sigma} \hat{n}_{k_\zeta,\sigma}$
is the occupation number operator for the electrons in each lead ($\zeta=L, R$).
For Hamiltonian (\ref{Htot}) the explicit 
expression for the current operator is given as
\begin{equation}
	\hat{I}_\zeta \equiv i[\hat{H}, \hat{N}_{\zeta}] = i \sum_{k_\zeta,\sigma}
	V_{dk_\zeta} ( \hat{d}^+_\sigma \hat{c}_{k_\zeta,\sigma} - \hat{c}_{k_\zeta,\sigma}^+ \hat{d}_\sigma ), \;\;\; 
	\zeta=L, R. 
\end{equation}

In the expressions above, $\hat{\rho}$ denotes the 
initial density matrix representing a grand-canonical ensemble for each lead and a certain
preparation for the bridge state
\begin{subequations}\label{Initden}
\begin{equation}
	\hat{\rho} = \hat{\rho}_d^0 \;{\rm exp} \left[ -\beta (\hat{H}_0 
          - \mu_L \hat{N}_L - \mu_R \hat{N}_R) \right],
\end{equation}
\begin{equation}
	\hat{H}_0 = \sum_{k_L,\sigma} E_{k_L} \hat{n}_{k_L,\sigma}
	+ \sum_{k_R,\sigma} E_{k_R} \hat{n}_{k_R,\sigma}  + \hat{H}_{\rm nuc}^0.
\end{equation}
\end{subequations}
Here $\hat{\rho}_d^0$ is the initial reduced density matrix for the bridge state, which is usually chosen as
a pure state representing an occupied or an empty bridge state, and $\hat{H}_{\rm nuc}^0$ defines the initial
bath equilibrium distribution. 

Various initial states can be considered.  For example, 
one may choose an initially unoccupied bridge state and the nuclear degrees of
freedom equilibrated with this state, i.e.\ an  unshifted 
bath of oscillators with $\hat H_{\rm nuc}^0$ 
as given in Eq.~(\ref{Hnuc}). On the other hand, one may also start with a fully occupied bridge state and 
a bath of oscillators in equilibrium with the occupied bridge state
\begin{equation}
{\hat H}_{\rm nuc}^{0'} = \frac{1}{2} \sum_j \left[ P_j^2 + \omega_j^2 \left(Q_j + 
\frac{c_j}{\omega_j^2}\right)^2 \right].
\end{equation}
Other initial states may also be prepared. The initial state may affect the
transient dynamics profoundly. The dependence of the steady-state current on
the initial density matrix is a more complex issue. Recent investigations for a model without
electron-electron interaction seem to indicate that different
initial states may lead to different (quasi)steady states,\cite{Gogolin02,Galperin05,Albrecht12} 
although this has been debated.\cite{Alexandrov07}
Even without coupling to a vibrational bath, the initial bridge state
population may still affect the final stationary current in a time-dependent
simulation.\cite{Dzhioev11,Khosravi12}
For all results reported in this paper, our calculations show that the stationary state is
independent on the initial condition within the error bar of the simulation
(which is estimated to be less than 10\% relative error). 
Since different sets of initial conditions also affect the time scale at which the current  $I(t)$ 
reaches its stationary value, we typically choose initial conditions that are close to the final steady 
state, e.g., an unoccupied initial bridge state if its energy is higher
than the Fermi level of the leads and an occupied bridge state otherwise.

The transient behavior of the thus defined currents $I_R(t)$ and $I_L(t)$ is usually different.  
However, the long-time limits of $I_R(t)$ and $I_L(t)$, which define the stationary current, are the same.
It is found that the average current 
\begin{equation}
	I(t) = \frac{1}{2} [ I_R(t) + I_L(t) ],
\end{equation} 
provides better numerical convergence properties by minimizing the transient characteristic, and 
thus will be used in most calculations.

In our simulations the continuous set of electronic states of the leads is represented by 
a finite number of states.  The number of states required to properly describe the 
continuum limit depends on the time $t$. The situation is thus similar to that of a quantum reactive 
scattering calculation in the presence of a scattering continuum, where, with a finite number of basis 
functions, an appropriate absorbing boundary condition is added to mimic the correct outgoing 
Green's function.\cite{Goldberg1978,Kosloff1986,Neuhauser1989,Seideman1991} 
Employing the same strategy for the present problem, the 
regularized electric current is given by
\begin{equation}
	I^{\rm reg}  = \lim_{\eta \to 0^+} \int_0^{\infty} dt \, \frac{dI(t)}{dt} \, e^{-\eta t}.
\end{equation}

The regularization parameter $\eta$ is similar (though not identical) to the formal convergence parameter
in the definition of  the Green's function in terms of the time evolution operator
\begin{equation}
	G(E^+) = \lim_{\eta \to 0^+} (-i) \int_0^{\infty} dt\, e^{i(E+i\eta-H)t}.
\end{equation}
In numerical calculations, $\eta$ is chosen in a similar way as the absorbing potential used in quantum
scattering calculations.\cite{Goldberg1978,Kosloff1986,Neuhauser1989,Seideman1991} In particular, 
the parameter $\eta$ has to be large enough to accelerate the convergence but still sufficiently small
in order not to affect the correct result.  While in the reactive scattering calculation
$\eta$ is often chosen to be coordinate dependent, in our simulation $\eta$ is chosen
to be time dependent
\begin{equation}\label{damping}
\eta(t) = \left\{
              \begin{array}{ll}
                   0 & \quad (t<\tau)\\
                   \eta_0\cdot (t-\tau)/t & \quad (t>\tau) .
              \end{array}
       \right.
\end{equation}
Here $\eta_0$ is a damping constant, $\tau$ is a cutoff time beyond which a steady state charge
flow is approximately reached.  As the number of electronic states increases, one may choose a 
weaker damping strength $\eta_0$ and/or longer cutoff time $\tau$.  The former approaches zero and the
latter approaches infinity for an infinite number of states.  In practice,
for the systems considered in this work, convergence can be reached with a reasonable number 
of electronic states in the range of 80-500, with a typical $\tau =$ 30-80 fs (a smaller
$\tau$ for less number of states) and $1/\eta_0 =$ 3-10 fs.

To gain insight into the transport mechanisms, it is also useful to consider the
population of the electronic states localized on the the molecular bridge,
which is given by
\begin{equation}\label{population}
P_d(t) = \frac{1}{{\rm tr}[\hat{\rho}]} {\rm tr}
        \left\{ \hat{\rho} e^{i\hat{H}t} \sum_{\sigma}\; \hat{n}_{d,\sigma}\; e^{-i\hat{H}t}
        \right\}.
\end{equation}

\section{The Multilayer Multiconfiguration Time-Dependent Hartree Theory
in Second Quantization Representation}\label{mlsqr}

The time-dependent study of transport properties in the model introduced above
requires a method that is able to 
describe many-body quantum dynamics in an accurate and efficient way.  For this purpose 
we employ the  recently proposed Multilayer Multiconfiguration Time-Dependent Hartree Theory
in Second Quantization Representation (ML-MCTDH-SQR),\cite{wan10:78} which
allows a numerically exact treatment of the many-body problem.  Here we give a brief
outline of the method.

\subsection{Overview of the ML-MCTDH theory}

The ML-MCTDH theory\cite{wan03:1289} is a rigorous variational method to propagate wave packets 
in complex systems with many degrees of freedom.  In this approach the wave function is represented 
by a recursive, layered expansion, 
\begin{subequations}\label{psiml}
\begin{equation}\label{L1}
        |\Psi (t) \rangle = \sum_{j_1} \sum_{j_2} ... \sum_{j_p}
        A_{j_1j_2...j_p}(t) \prod_{\kappa=1}^{p}  |\varphi_{j_\kappa}^{(\kappa)} (t) \rangle,
\end{equation}
\begin{equation}\label{L2}
        |\varphi_{j_\kappa}^{(\kappa)}(t)\rangle =  \sum_{i_1} \sum_{i_2} ... \sum_{i_{Q(\kappa)}}
        B_{i_1i_2...i_{Q(\kappa)}}^{\kappa,j_\kappa}(t) \prod_{q=1}^{Q(\kappa)}  
	|v_{i_q}^{(\kappa,q)}(t) \rangle,
\end{equation}
\begin{equation}\label{L3}
        |v_{i_q}^{(\kappa,q)}(t)\rangle  = \sum_{\alpha_1} \sum_{\alpha_2} ... 
	\sum_{\alpha_{M(\kappa,q)}}
        C_{\alpha_1\alpha_2...\alpha_{M(\kappa,q)}}^{\kappa,q,i_q}(t) 
	\prod_{\gamma=1}^{M(\kappa,q)}  
	|\xi_{\alpha_\gamma}^{\kappa,q,\gamma}(t) \rangle,
\end{equation}
\begin{equation}
	... \nonumber
\end{equation}
\end{subequations}
where $A_{j_1j_2...j_p}(t)$, $B_{i_1i_2...i_{Q(\kappa)}}^{\kappa,j_\kappa}(t)$,
$C_{\alpha_1\alpha_2...\alpha_{M(\kappa,q)}}^{\kappa,q,i_q}(t)$ and so on are the
expansion coefficients for the first, second, third, ..., layers, respectively;
$|\varphi_{j_\kappa}^{(\kappa)} (t) \rangle$, $|v_{i_q}^{(\kappa,q)}(t) \rangle$,
$|\xi_{\alpha_\gamma}^{\kappa,q,\gamma}(t) \rangle$, ..., are the ``single particle'' 
functions (SPFs) for the first, second, third, ..., layers.  
In Eq.~(\ref{L1}), $p$ denotes the number of single
particle (SP) groups/subspaces for the first layer.  Similarly, $Q(\kappa)$ in Eq.~(\ref{L2})
is the number of SP groups for the second layer that belongs to the $\kappa$th SP
group in the first layer,  i.e., there are a total of $\sum_{\kappa=1}^{p} Q(\kappa)$
second layer SP groups.  Continuing along the multilayer hierarchy, 
$M(\kappa,q)$ in Eq.~(\ref{L3}) is the number of SP groups for the third layer that belongs 
to the $q$th SP group of the second layer and the $\kappa$th SP group of the first layer,  
resulting in a total of $\sum_{\kappa=1}^{p} \sum_{q=1}^{Q(\kappa)} M(\kappa,q)$ third 
layer SP groups.  Naturally, the size of the  system that the ML-MCTDH theory can treat 
increases with the number of layers in the expansion.  In principle, such a recursive 
expansion can be carried out to an arbitrary number of layers.  The multilayer hierarchy 
is terminated at a particular level by expanding the SPFs in the deepest layer in terms of 
time-independent configurations, each of which may contain several Cartesian degrees of 
freedom.

The variational parameters within the ML-MCTDH theoretical framework are dynamically 
optimized through the use of the Dirac-Frenkel variational principle\cite{Frenkel34}
\begin{equation}
        \langle \delta\Psi(t) | i \frac{\partial} {\partial t} - \hat{H} |
        \Psi(t) \rangle = 0,
\end{equation}
which results in a set of coupled, nonlinear differential equations for the expansion
coefficients for all layers.\cite{wan03:1289,wan10:78,wan09:024114} For a $N$-layer version 
of the ML-MCTDH theory there are $N+1$ levels of expansion coefficients.  In this sense the 
conventional wave packet propagation method is a ``zero-layer'' MCTDH approach.

The introduction of this recursive, dynamically optimized layering scheme in the ML-MCTDH 
wavefunction provides more flexibility in the variational functional, which results in 
a tremendous gain in our ability to study large  many-body quantum systems.  During the past 
few years, significant progress has been made in further development of the theory to simulate 
quantum dynamics and nonlinear spectroscopy of ultrafast  electron transfer reactions in 
condensed phases.\cite{tho06:210,wan06:034114,kon06:1364,wan07:10369,kon07:11970,tho07:153313,wan08:139,wan08:115005,ego08:214303,vel09:094109,wan10:78,Vendrell11} 
The theory has also been generalized to study heat transport in molecular 
junctions\cite{vel08:325} and to calculate rate constants for model proton transfer reactions 
in molecules in solution.\cite{wan06:174502,cra07:144503}
Recent work of Manthe has introduced an even more adaptive formulation based on
a layered correlation discrete variable representation (CDVR).\cite{man08:164116,man09:054109}

\subsection{Treating identical particles using the second quantization representation of Fock space}

To extend the original ML-MCTDH approach to systems of identical quantum
particles requires a method that incorporates the exchange symmetry explicitly.  
This is because an ordinary Hartree product 
within the first quantized picture is only suitable to describe a configuration for a system of 
distinguishable particles.  One strategy is to employ a properly symmetrized wave function in
the first quantized framework, i.e., permanents in a bosonic case or Slater determinants in a 
fermionic case.  This led to the MCTDHF approach\cite{kat04:533,cai05:012712,nes05:124102} for 
treating identical fermions and the MCTDHB approach\cite{alo08:033613} for treating identical bosons 
as well as combinations thereof.\cite{alo07:154103}  However, this wave function-based symmetrization 
is only applicable to the single layer MCTDH theory but is incompatible with the ML-MCTDH theory 
with more layers --- there is no obvious analog of a multilayer Hartree configuration if 
permanents/determinants are used to represent the wave function.  As a result, the ability to treat 
much larger quantum systems numerically exactly was severely limited.

To overcome this limitation we proposed a novel approach\cite{wan09:024114} that follows a 
fundamentally different route to tackle many-body quantum dynamics of indistinguishable particles --- 
an operator-based method that employs the second quantization formalism of many-particle quantum 
theory.  This differs from many previous methods where the second quantization formalism is only 
used as a convenient tool to derive intermediate expressions for the first quantized form.  In the 
new approach the variation is carried out entirely in the abstract Fock space using the occupation 
number representation. Therefore, the burden of handling symmetries of identical particles in a 
numerical variational calculation is shifted completely from wave functions to the algebraic properties 
of operators.

The major difference between the ML-MCTDH-SQR theory for identical fermions and the previous 
ML-MCTDH theory for distinguishable particles is the way how operators act. For example, in the second 
quantized form the fermionic creation/annihilation operators fulfill the 
anti-commutation relations
\begin{equation}\label{anticomm}
	\{ \hat{a}_P, \hat{a}_Q^+ \} \equiv \hat{a}_P \hat{a}_Q^+ + \hat{a}_Q^+ \hat{a}_P = \delta_{PQ},
	\hspace{1cm}
	\{ \hat{a}_P^+, \hat{a}_Q^+ \} = \{ \hat{a}_P, \hat{a}_Q \} = 0.
\end{equation}
The symmetry of identical particles is thus realized by enforcing such algebraic
properties of the operators.  This can be accomplished by introducing a permutation
sign operator associated with each fermionic creation/annihilation operator, which incorporates 
the sign changes of the remaining spin orbitals in all the SPFs whose subspaces are prior to 
it.\cite{wan09:024114} For example, if a purely electronic problem is
considered and only one layer is present, the overall wave 
function and the SPFs have the form
\begin{subequations}
\begin{equation}
\label{mcsqr}
        |\Psi (t) \rangle 
	= \sum_{j_1} \sum_{j_2} ... \sum_{j_L}
        A_{j_1j_2...j_L}(t) \prod_{\kappa=1}^{L}  |\varphi_{j_\kappa}^{(\kappa)} (t) \rangle,
\end{equation}
\begin{equation}\label{spsqr}
        |\varphi_{j_\kappa}^{(\kappa)}(t)\rangle = \sum_{I_\kappa=1}^{2^{m_\kappa}}
	B_{I_\kappa}^{\kappa,j_\kappa}(t) |\phi^{(\kappa)}_{I_\kappa} \rangle \equiv
	\sum_{n_1=0}^1 \sum_{n_2=0}^1 ... \sum_{n_{m_\kappa}=0}^1
        B_{n_1n_2...n_{m_\kappa}}^{\kappa,j_\kappa}(t)\; |n_1\rangle |n_2\rangle ...  
	|n_{m_\kappa} \rangle,
\end{equation}
\end{subequations}
where $n_i=0,1$ are the occupation numbers.  A fermionic creation operator is actually
implemented as
\begin{equation}\label{fermcreat}
	({a}_{\nu}^{(\kappa)})^+ = \left( \prod_{\mu=1}^{\kappa-1}\; \hat{S}_\mu \right)
	\; ({\tilde{a}}_{\nu}^{(\kappa)})^+,
\end{equation}
where $\hat{S}_\mu$ ($\mu=1,2,...,\kappa-1$) is the permutation sign operator that accounts 
for permuting $({a}_{\nu}^{(\kappa)})^+$ from the first subspace all the way through to 
the $\kappa$th subspace, and
$({\tilde{a}}_{\nu}^{(\kappa)})^+$ is the reduced creation operator that only takes care of 
the fermionic anti-commutation relation in the $\kappa$th subspace.  The operator-based
anti-commutation constraint (\ref{anticomm}) results in the following operations
\begin{subequations}\label{permutesign}
\begin{equation}
	({\tilde{a}}_{\nu}^{(\kappa)})^+  |\varphi_{j_\kappa}^{(\kappa)}(t)\rangle =
	\sum_{n_1=0}^1 \sum_{n_2=0}^1 ... \sum_{n_{m_\kappa}=0}^1 \; \delta_{n_\nu,0}
	\left[\prod_{q=1}^{\nu-1} (-1)^{n_q}\right]\;  B_{n_1n_2...n_{m_\kappa}}^{\kappa,j_\kappa}(t)\;
	|n_1\rangle |n_2\rangle ... |1_\nu\rangle ... |n_{m_\kappa} \rangle,
\end{equation}
\begin{equation}
	\hat{S}_\mu |\varphi_{j_\mu}^{(\mu)}(t)\rangle =
	\sum_{n_1=0}^1 \sum_{n_2=0}^1 ... \sum_{n_{m_\mu}=0}^1 \;
	\left[\prod_{q=1}^{m_\mu} (-1)^{n_q}\right]\;  B_{n_1n_2...n_{m_\mu}}^{\mu,j_\mu}(t)\;
	|n_1\rangle |n_2\rangle ... |n_{m_\mu} \rangle.
\end{equation}
\end{subequations}
I.e.,  $({\tilde{a}}_{\nu}^{(\kappa)})^+$ not only creates a particle in the 
$\nu$th spin orbital if it is vacant, but also affects the sign of each term in 
this SPF according to where 
$\nu$ is located and what the occupations are prior to it.  Furthermore, the permutation sign 
operators  $\hat{S}_\mu$, $\mu=1,2,...,\kappa-1$, incorporate the sign changes of the remaining 
spin orbitals in all the SPFs whose subspaces are prior to that of 
$(\tilde{a}_{\nu}^{(\kappa)})^+$.

Thus, the occupation number states in the ML-MCTDH-SQR theory are treated in the same way as the degrees 
of freedom in the original ML-MCTDH theory, except that the orderings of all the SP groups in all 
layers need to be recorded and maintained in later manipulations.  More importantly, the equations 
of motion have the same form as in the original ML-MCTDH theory.  The only difference is that 
for identical fermions each creation/annihilation operator of the 
Hamiltonian is effectively a product of operators: a reduced creation/annihilation operator 
that only acts on the bottom-layer SPFs for the Fock subspace it belongs to, and a series of 
permutation sign operators that accounts for the fermionic anti-commutation relations of all 
the spin orbitals prior to it.
In the multilayer case the implementation is sophisticated but can still be reduced to handling
(many and complicated) basic building blocks in the MCTDH or ML-MCTDH theory --- products of operators.  
Thereby, the action of each Hamiltonian term (product of creation/annihilation operators) can be 
split into a series of operations on individual Fock subspaces.\cite{wan09:024114}  
On the other hand, for identical bosons the implementation
is much simpler because there is no sign change upon permutation.

In the second quantized form,  the wave function is represented in the abstract Fock space 
employing the occupation number basis.  As a result, it can be expanded in the same multilayer 
form as that for systems of distinguishable particles.  It is thus possible to extend the 
numerically exact treatment to much larger systems.  The symmetry of the wave function in 
the first quantized form is shifted to the operator algebra in the second quantized form.  
The key point is that, for both phenomenological models and more fundamental theories, 
there are only a limited number of combination of fundamental operators.  
For example, in electronic structure theory only one- and two-electron operators are present.  
This means that one never needs to handle all, redundant possibilities of operator combinations as 
offered by the determinant form in the first quantized framework.  It is exactly this property 
that provides the flexibility of representing the wave functions in multilayer form and treat 
them accurately and efficiently within the ML-MCTDH-SQR theory. It is also noted that 
the ML-MCTDH-SQR approach outlined above for fermions has
also be formulated for bosons or combinations of fermions, bosons and
distinguishable particles.\cite{wan09:024114}

\section{Results and Discussion}\label{results}

In this section, we present applications of the ML-MCTDH-SQR methodology to
the study of correlated electron transport employing the model described in
Sec.~\ref{modeltight}. In particular, we discuss the influence of
electron-electron and electronic-vibrational interaction on the transport
characteristics for selected examples. Unlike the noninteracting transport
model ($U_d = 0, c_j = 0$), these results represent nontrivial
solutions to a many-body quantum problem and are often beyond the perturbation
treatment. All calculations presented in this paper are for zero temperature,
which corresponds to the deep quantum regime, and is often the most challenging
physical regime of the problem.  Meanwhile, this regime is relatively easy 
for our approach since only one initial wave function is required.
An investigation of systems at finite temperature as well as an analysis of
the physical mechanisms in a broader
parameter range will be the topic of future work.

\subsection{Effect of electron-electron interaction on transport characteristics for fixed nuclei}

\begin{figure}[!ht]
\begin{flushleft}
(a)
\end{flushleft}
\includegraphics[clip,width=0.45\textwidth]{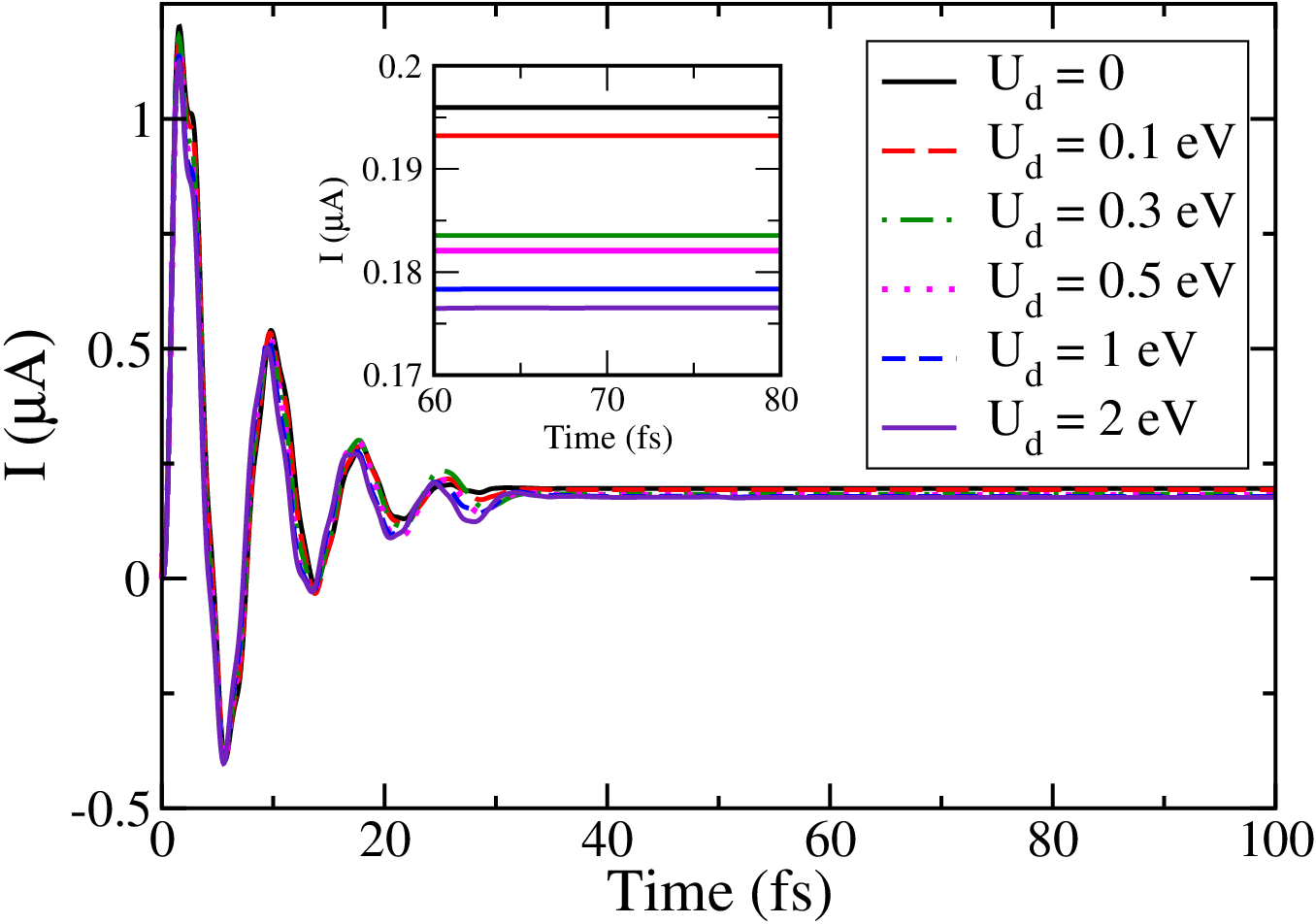}

\begin{flushleft}
(b)
\end{flushleft}
\includegraphics[clip,width=0.45\textwidth]{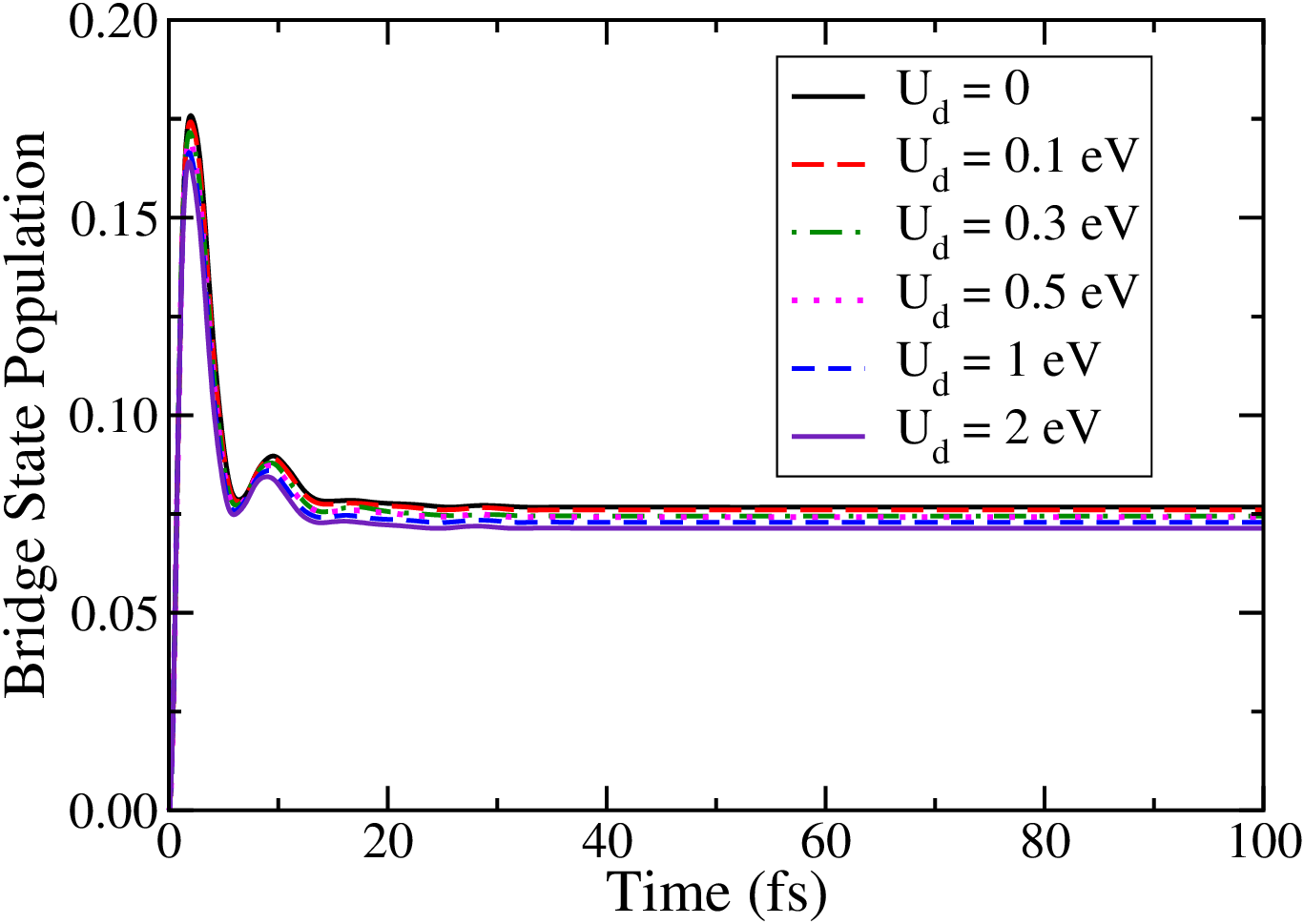}

\caption{(a) Time-dependent current $I(t)$ for different electron-electron
  coupling strength $U_d$ and (b)
the corresponding electronic population at the bridge state.
Other parameters are: $\alpha_e = 0.2$eV, $\beta_e = 1$eV, $E_d - E_f = 0.5$eV,
and the source-drain voltage $V = 0.1$V. The inset in panel (a) depicts the
stationary current in an enlarged view.}
\label{fig1}
\end{figure}

We first focus on the influence of electron-electron interaction and
consider models without electronic-vibrational coupling ($c_j=0$),
i.e.\ for fixed nuclei. 
Fig.~\ref{fig1} shows the time-dependent current and the corresponding
bridge state population for a model with the following set of electronic
parameters:  The tight-binding parameters for the 
function $\Gamma (E)$ are $\alpha_e = 0.2$ eV, $\beta_e = 1$ eV, 
corresponding to a moderate molecule-lead coupling and a bandwidth of 4 eV.
The energy of the bridge  states $E_d$ is located 0.5 eV above Fermi energy and the 
source-drain voltage is $V=0.1$ V, i.e.\ the model is in the off-resonant
transport regime. The results for both the current and the population show
pronounced transient oscillations that decay on a time scale 
of $\approx \Gamma^{-1}$ and approach a stationary plateau at longer times,
which  represents the steady state. The overall values of the current and population are
rather small because the transport takes place in the off-resonant regime. 
The comparison of the results obtained for different parameters $U_d$
shows that for this model electron-electron
interaction has no significant influence on the population and the current,  
and this includes both the transient behavior and the long-time stationary value.
Qualitatively this can be understood from the fact that the model is in the off-resonant transport 
regime. At zero coupling strength $U_d$, the bare energies of the electronic 
bridge states are the same and are outside the conductance window defined by the chemical potentials 
of the two electrodes.  Including the on-site repulsion term removes the degeneracy of these
two bridge states if one state is occupied.  That is, when one of the bridge states is populated, 
the electronic energy of the other state is increased by the value of $U_d$.  However, due to the fact 
that the initial bridge states are relatively far away from the conductance window, their populations 
are small.  As a result, the overall electronic correlation effect is small for this 
set of model parameters. At a finer scale
it can be seen that with the increase in $U_d$ both the 
stationary current and bridge state population decreases.  This is consistent with the fact that upon 
increasing $U_d$ the energy of the doubly occupied state ($E_d + U_d$) is moved to higher
energies and thus even further away from the conduction window.

\begin{figure}[!ht]
\begin{flushleft}
(a)
\end{flushleft}
\includegraphics[clip,width=0.45\textwidth]{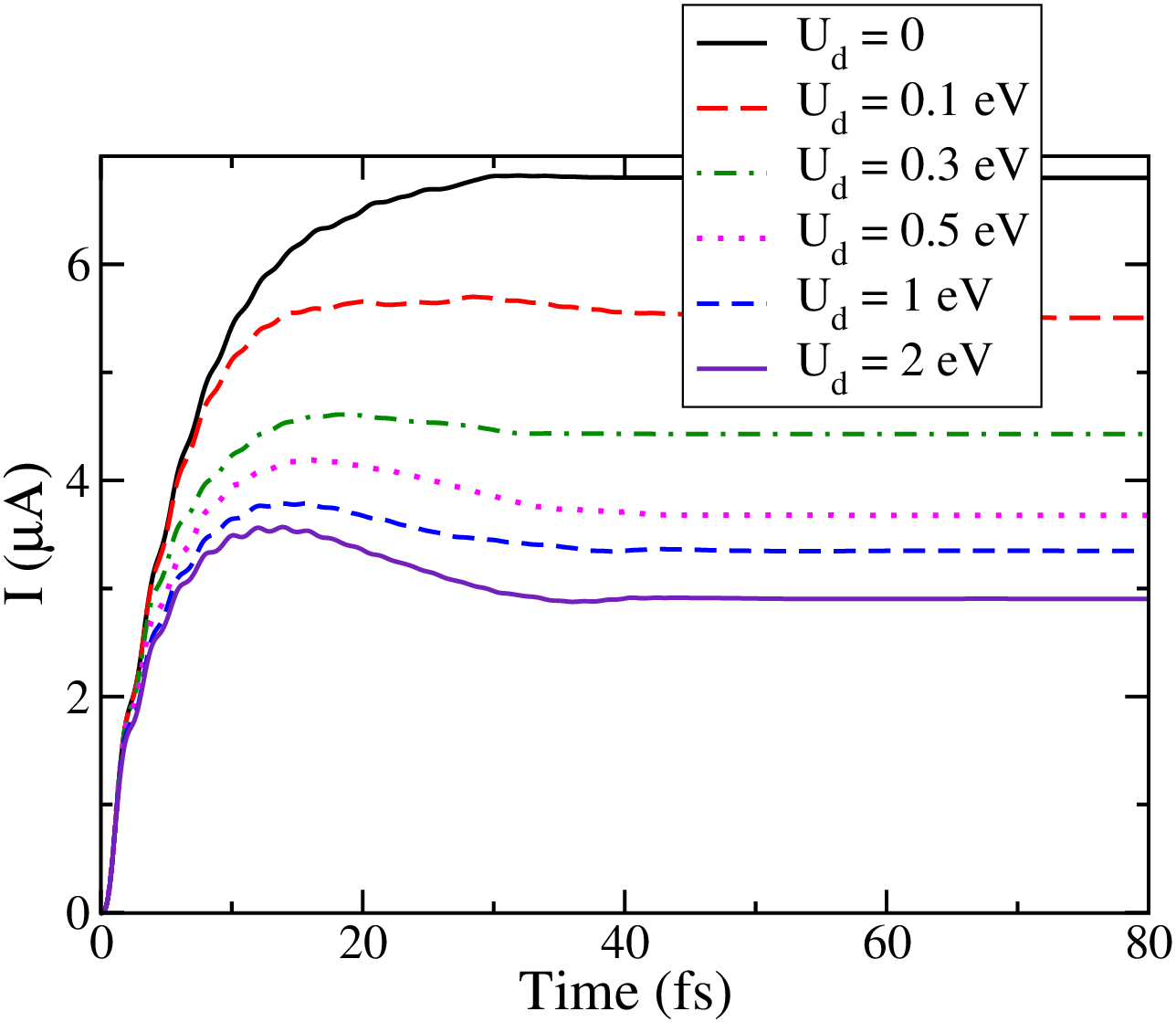}

\begin{flushleft}
(b)
\end{flushleft}
\includegraphics[clip,width=0.45\textwidth]{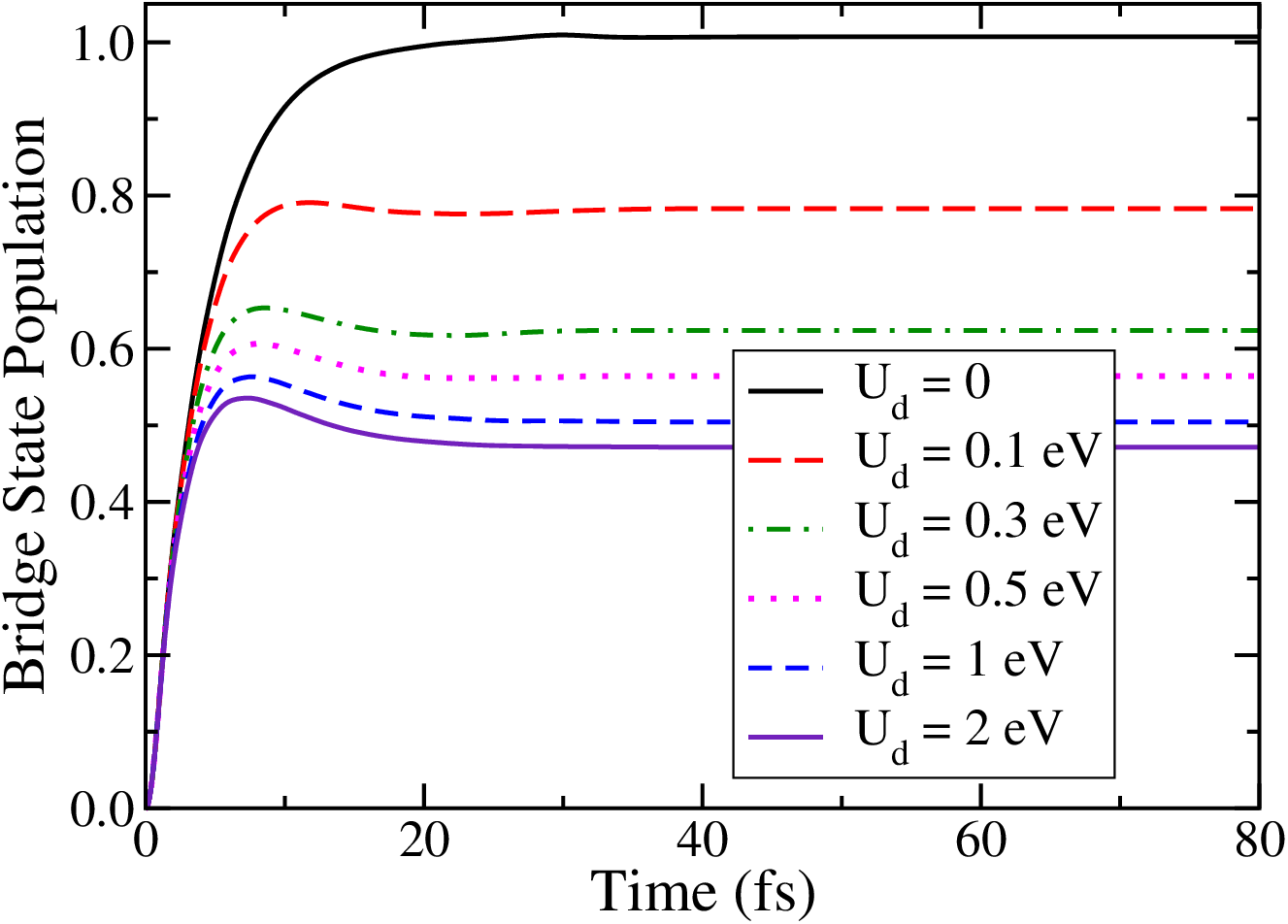}

\caption{Same as Fig.~\ref{fig1} except for $E_d - E_f = 0$.}
\label{fig2}
\end{figure}

Figure~\ref{fig2} shows the time-dependent current and the corresponding bridge state population for
another model, where the parameters are the same as in Fig.~\ref{fig1} except for $E_d-E_f=0$, i.e.,
energy of the bridge states is located at the Fermi energy of the leads.
For zero on-site coupling strength ($U_d=0$), this set of parameters
corresponds to the resonant tunneling regime and involves
mixed electron/hole transport. This results in a significantly larger
stationary current and a population of approximately one, because each bridge
state has ~50\% probability to be occupied.
In this parameter regime, electron-electron interaction has a pronounced
influence on the transport characteristics. Upon increase of $U_d$ the steady
state value of both the current and the population decreases
significantly. This is due to the fact that for increasing $U_d$ the energy of
the doubly occupied state moves out of the conductance window. For interaction strengths
$U_d\gg 0.1$eV, the bridge state can only be singly occupied, resulting in an overall
population of $n_d = 1/2$, and the doubly occupied state does not contribute to
the current. 

\begin{figure}[!ht]
\begin{flushleft}
(a)
\end{flushleft}
\includegraphics[clip,width=0.45\textwidth]{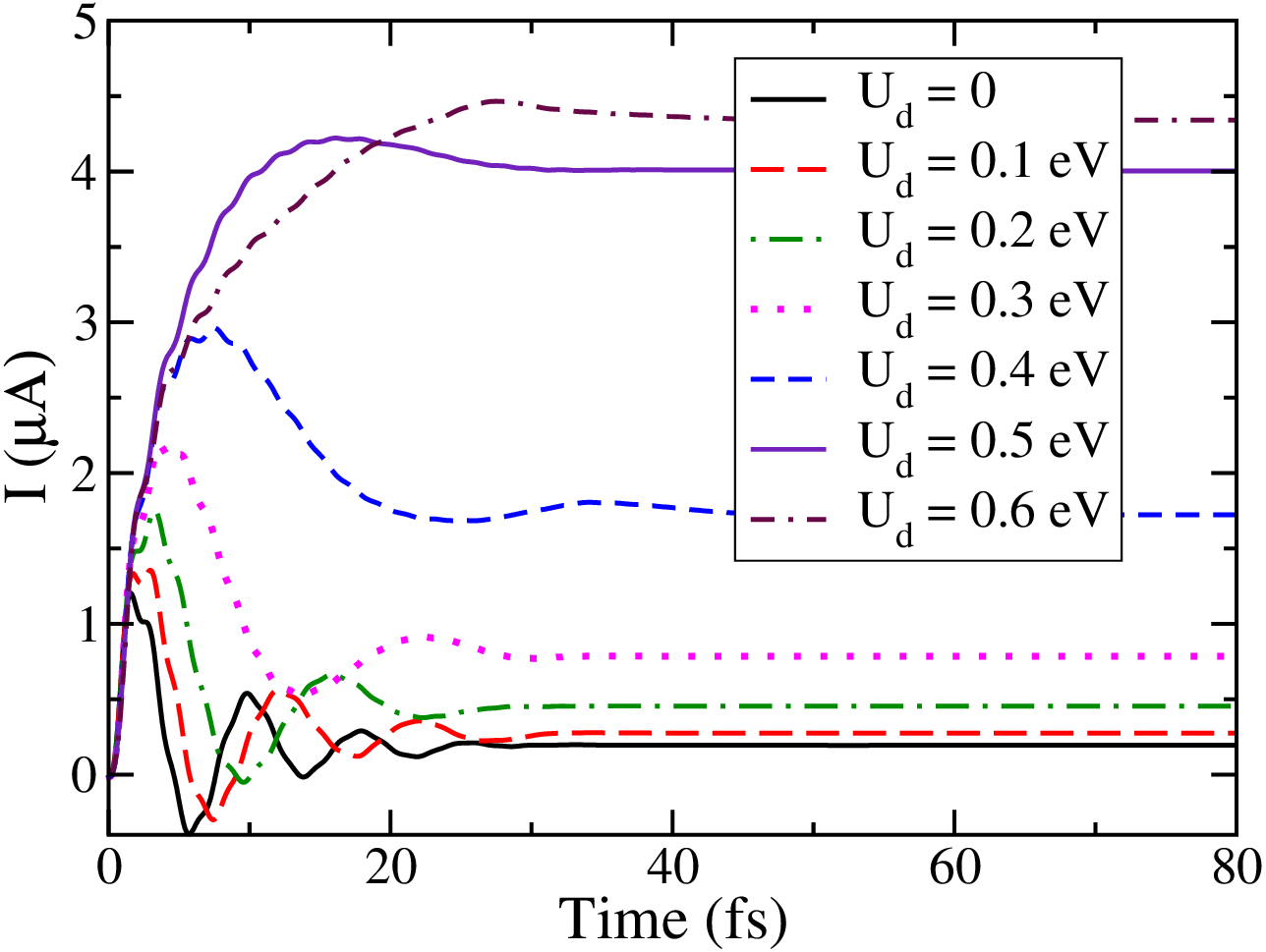}

\begin{flushleft}
(b)
\end{flushleft}
\includegraphics[clip,width=0.45\textwidth]{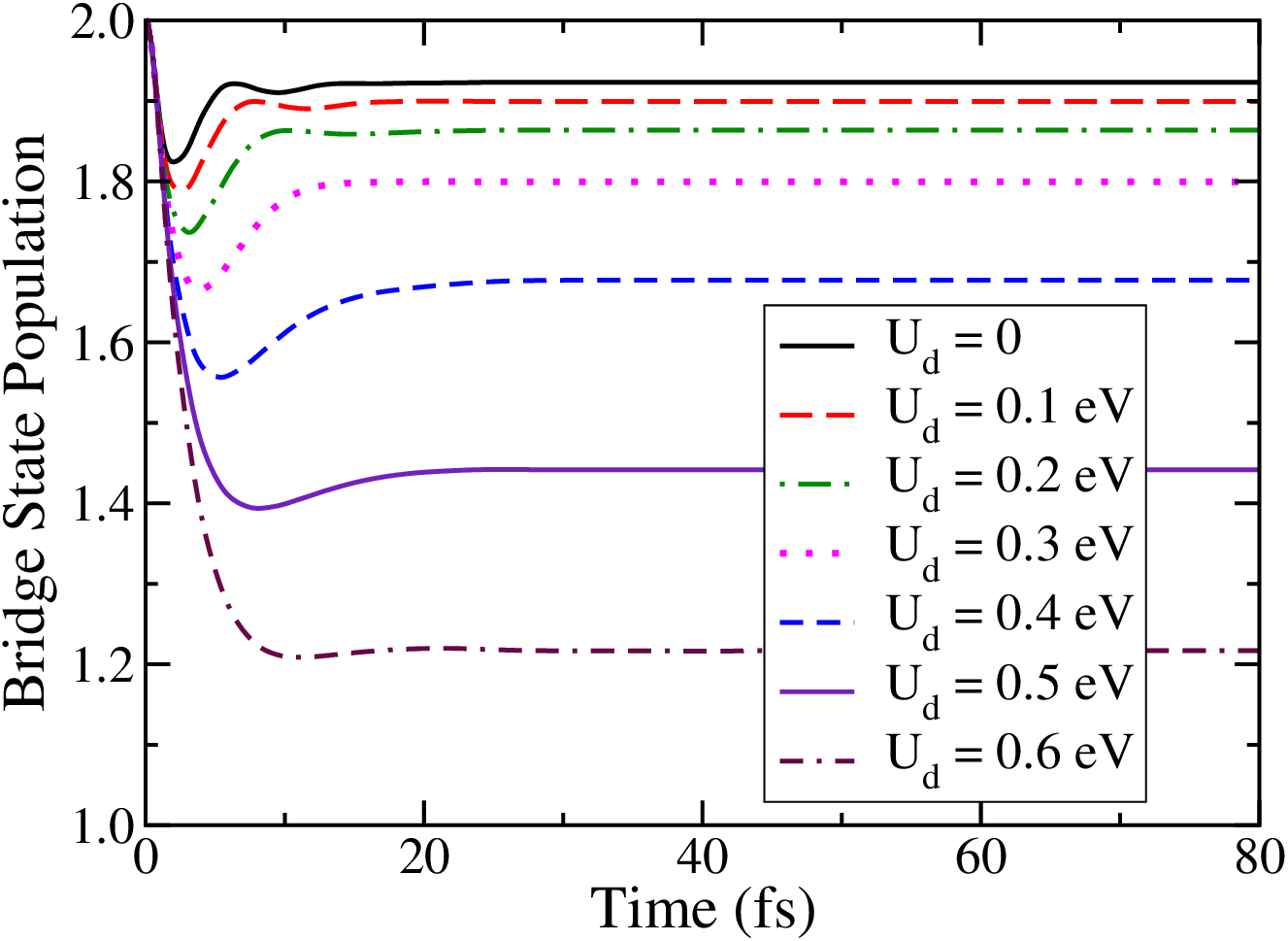}

\caption{Same as Fig.~\ref{fig1} except for $E_d - E_f = -0.5$eV.
}
\label{fig3}
\end{figure}

\begin{figure}[!ht]
\begin{flushleft}
(a)
\end{flushleft}
\includegraphics[clip,width=0.45\textwidth]{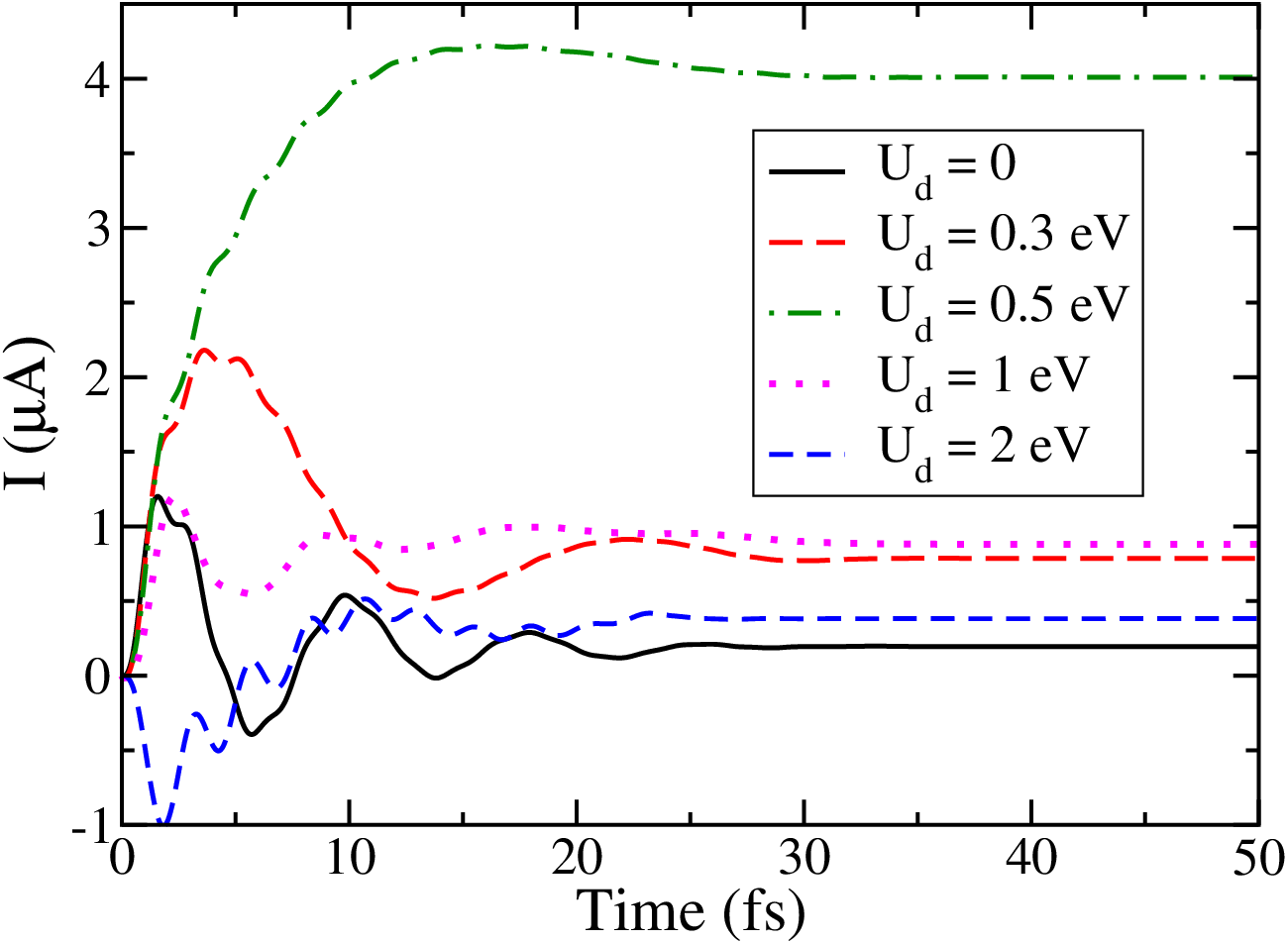}

\begin{flushleft}
(b)
\end{flushleft}
\includegraphics[clip,width=0.45\textwidth]{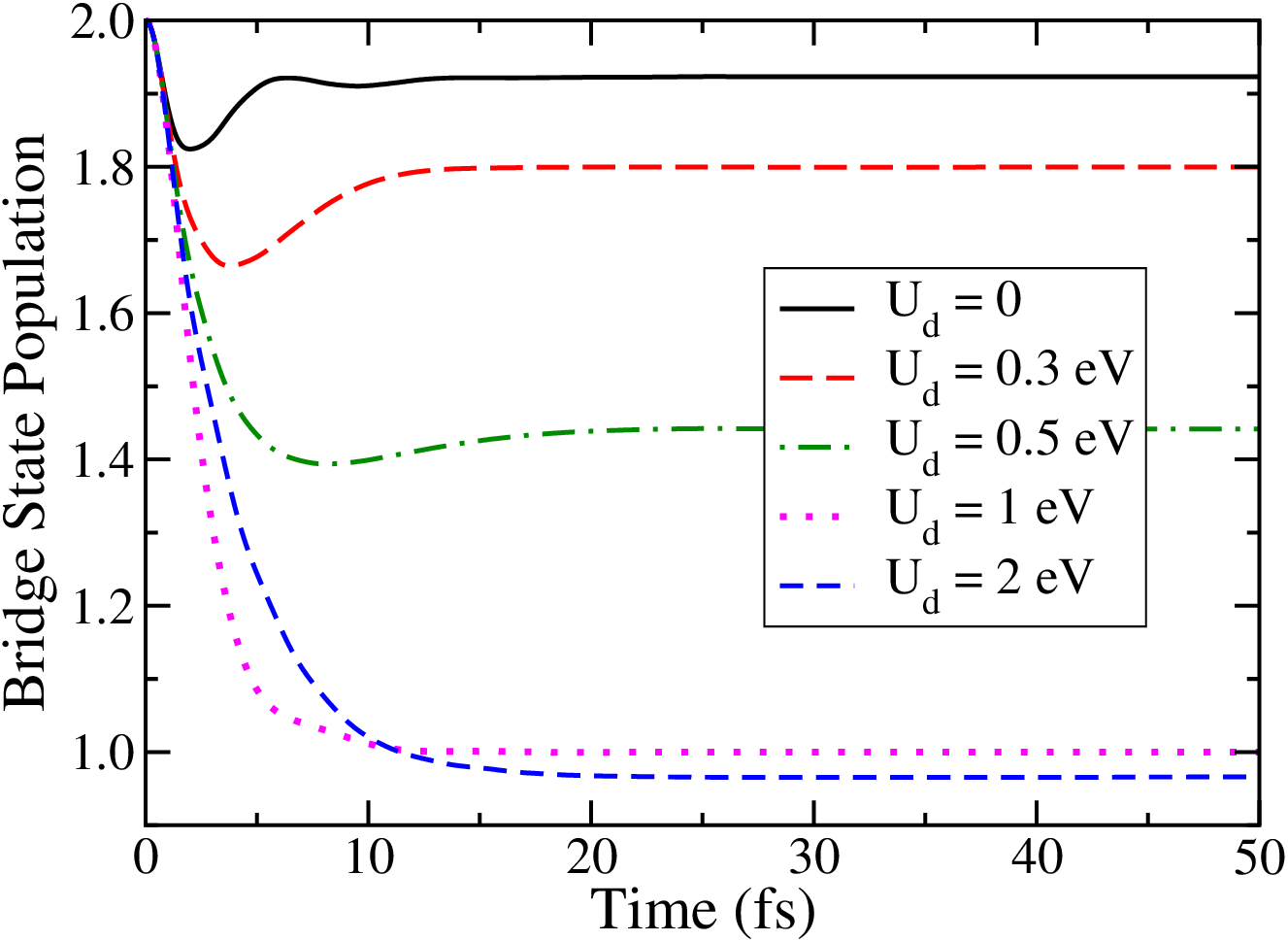}

\caption{Same as Fig.~\ref{fig3} except for a larger range of $U_d$.}
\label{fig4}
\end{figure}

We next consider in Figs.~\ref{fig3}, \ref{fig4} a model with the same
parameters as in the previous 
two cases except that the energy 
of the bridge state is below the Fermi energy, $E_d - E_f= -0.5$ eV. For
vanishing electron-electron interaction, $U_d=0$, this is again a
non-resonant case. However, because  the bridge state is, in contrast to Fig.~\ref{fig1}, located
below the Fermi energy it is almost doubly
occupied when $U_d=0$. While the stationary current for $U_d=0$ (full black line in
Fig.~\ref{fig3}(a)) is, due to particle-hole
symmetry, in fact identical to that in Fig.~\ref{fig1}(a), the dependence of
the transport characteristics on the electron-electron interaction is more
complex than in the above two cases. Upon moderate increase of $U_d$, the energy of
the doubly occupied state, $E_d + U_d$, moves closer to the Fermi energy and
enters the conductance window. As a result, the population of this state
decreases as shown in Fig.~\ref{fig3}(b). The current (Fig.~\ref{fig3}(a)), on the other hand, increases because the doubly
occupied state provides a channel for resonant transport. It is also observed that
upon moderate increase of $U_d$ the transient dynamics undergoes a coherent to incoherent transition. 
When $U_d$ is further increased ($U_d > 0.6$ eV), the
energy of the doubly occupied state becomes higher than the chemical potentials
of both electrodes. As shown in Fig.~\ref{fig4}, this causes a decrease of the
current and the population. For large values of $U_d$, the population in the
steady state approaches a  value of unity, because the bridge state can only be
singly occupied.
It is interesting to note that for large Coulomb repulsion, e.g.\  $U_d=2$eV, the initial transient current
is negative.  This is because in the simulation the bridge state is initially fully occupied.  During the
early transient time, electrons flow from the bridge states to both the left and the right electrodes,
resulting in a negative transient current.  As the bridge states approach their steady state population,
electrons move continuously from the left electrode to the right electrode, establishing a steady-state
current.

\subsection{Aspects of Coulomb blockade}

An interesting  many-body
nonequilibrium effect in charge transport in mesoscopic and nanosystems is 
Coulomb blockade.\cite{Glazman89,Beenakker91,Grabert92}  This phenomenon involves the suppression of the electrical 
current due to electron-electron interaction. Within the single-site Anderson
impurity model, the underlying mechanism is that the Coulomb repulsion
with an electron that already occupies the bridge state prevents a second
electron to transfer onto the bridge and thus reduces the current compared to
a noninteracting model.

\begin{figure}[!ht]

\includegraphics[clip,width=0.45\textwidth]{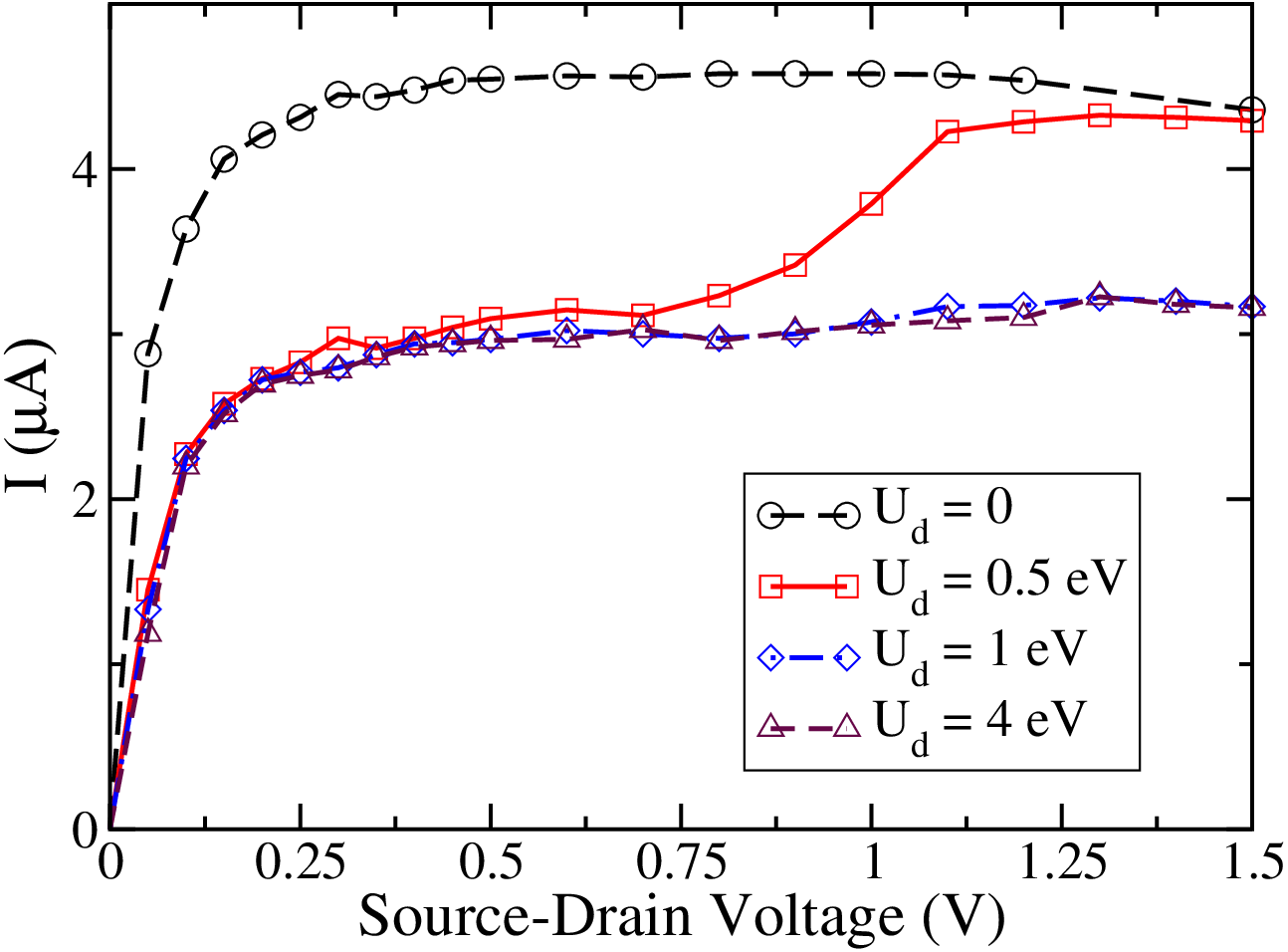}

\caption{Current-voltage characteristics for different electron-electron
  coupling strength $U_d$. Other parameters are: $\alpha_e = 0.1$eV, $\beta_e = 1$eV, $E_d - E_f = 0$.
The lines are intended as a guide to the eye.}
\label{fig5}
\end{figure}

This basic aspect of Coulomb blockade is demonstrated in Fig.~\ref{fig5},
which shows simulated current-voltage characteristics for a resonant transport
model, where the energy of the bridge states $E_d$ is at the Fermi energy of
the leads, $E_d-E_f=0$. The tight-binding parameters 
for the function $\Gamma (E)$ are $\alpha_e = 0.1$ eV, $\beta_e = 1$ eV, corresponding to a smaller 
molecule-lead coupling and a bandwidth of 4 eV. Besides the noninteracting model ($U_d=0$),
three values of the electron-electron coupling strengths are considered:
$U_d=0.5,$ 1, and 4 eV. To obtain the current-voltage characteristics, the
stationary plateau value from the time-dependent simulation of the current was
taken for each given voltage.
The results show that upon inclusion of electron-electron interaction, 
the currents are suppressed at all voltages. The ratio
between the blocked and unblocked currents attain a stationary ratio of approximately 2/3 in the 
plateau region (within the convergence range of less than 10\% relative error), and is nearly 
independent of the electronic coupling strength $U_d$. Within a zeroth order picture, this result 
can be rationalized as follows.

For the model without electron-electron interaction ($U_d=0$), there are three channels for electron
transport through the two bridge states: (i) Electron transport through an
initially unoccupied state.  There are two such channels
corresponding to the two spin polarizations of the bridge state.
(ii) Electron transport through an initially singly occupied bridge resulting
in the third channel which involves double occupation of the bridge state.
When the source-drain voltage $V$ is small, roughly $|eV|< 2U_d$ in the zeroth-order picture, the
third two-electron transport channel is essentially closed, resulting in a
current value 2/3 of that for 
the unblocked case.  At approximately $|eV|= 2U_d$, e.g., $V=1$ V for the case $U_d=0.5$ eV in Fig.~\ref{fig5},
the two-electron transport channel becomes available and the current begins to increase with the
source-drain voltage.  For finite molecule-lead coupling, the transition is broadened as shown in
Fig.~\ref{fig5}. For larger values of $U_d$, the energy of the doubly occupied
state is outside the conductance window of the bias voltages considered and thus the current is suppressed.

\begin{figure}[!ht]

\includegraphics[clip,width=0.45\textwidth]{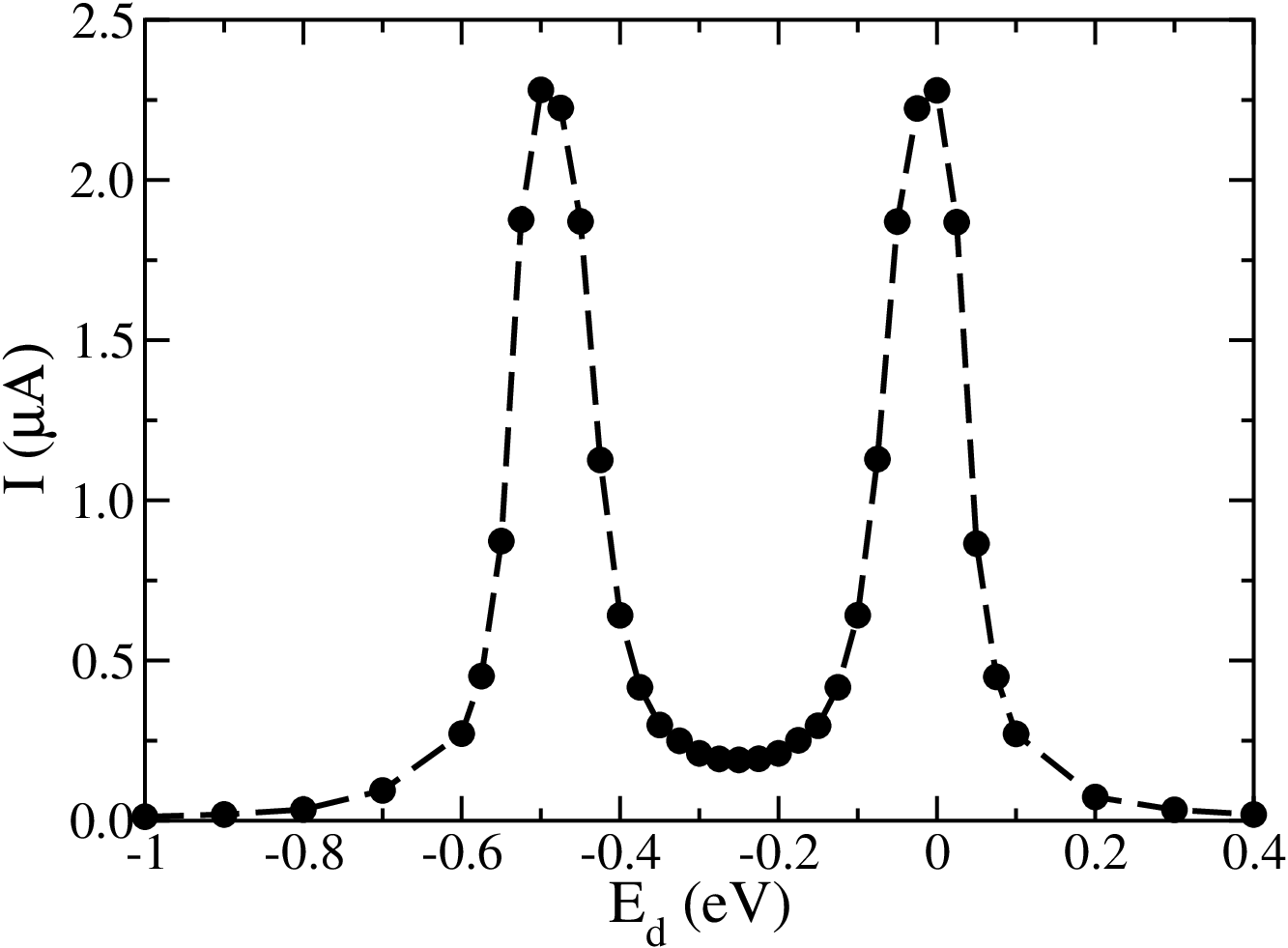}

\caption{Dependence of the current on the gate voltage.  The electronic parameters are: 
$\alpha_e = 0.1$eV, $\beta_e = 1$eV, $U_d = 0.5$eV, and $E_d - E_f = 0$ for zero gate
voltage.  The source-drain voltage is 0.1V. The line is intended as a guide to the eye.}
\label{fig6}
\end{figure}

While Fig.~\ref{fig5} demonstrates the phenomenon of Coulomb blockade for
varying the source-drain voltage, it is also instructive to study the phenomenon
for varying the gate voltage. Assuming that an additional gate voltage  $V_g$
predominantly shifts the energy of the bridge states $E_d$, we can investigate the
influence of the gate voltage by varying $E_d$ relative to the Fermi energy of
the leads. The result depicted in Fig.~\ref{fig6}  exhibits the well known peak
structures of Coulomb blockade, with maxima at
energies $E_d=0$ and $E_d=-U_d$, where the singly occupied levels and the
doubly occupied level are in the conductance window, respectively.
The   parameters here are the same as in Fig.~\ref{fig5} except for a fixed
$U_d=0.5$eV and a source-drain voltage $V = 0.1$V. It is noted that
the value of the voltage considered is already beyond the linear response regime.

\subsection{Effect of electron-electron interaction on transport characteristics in the presence of electron-vibrational coupling}

We finally consider a model which includes both electron-electron and electron-vibrational
interaction. The presence of both interactions increases the complexity
significantly. 
To the best of our knowledge, the results presented here are the first
numerical exact simulations for this type of models.

\begin{figure}[!ht]
\begin{flushleft}
(a)
\end{flushleft}
\includegraphics[clip,width=0.45\textwidth]{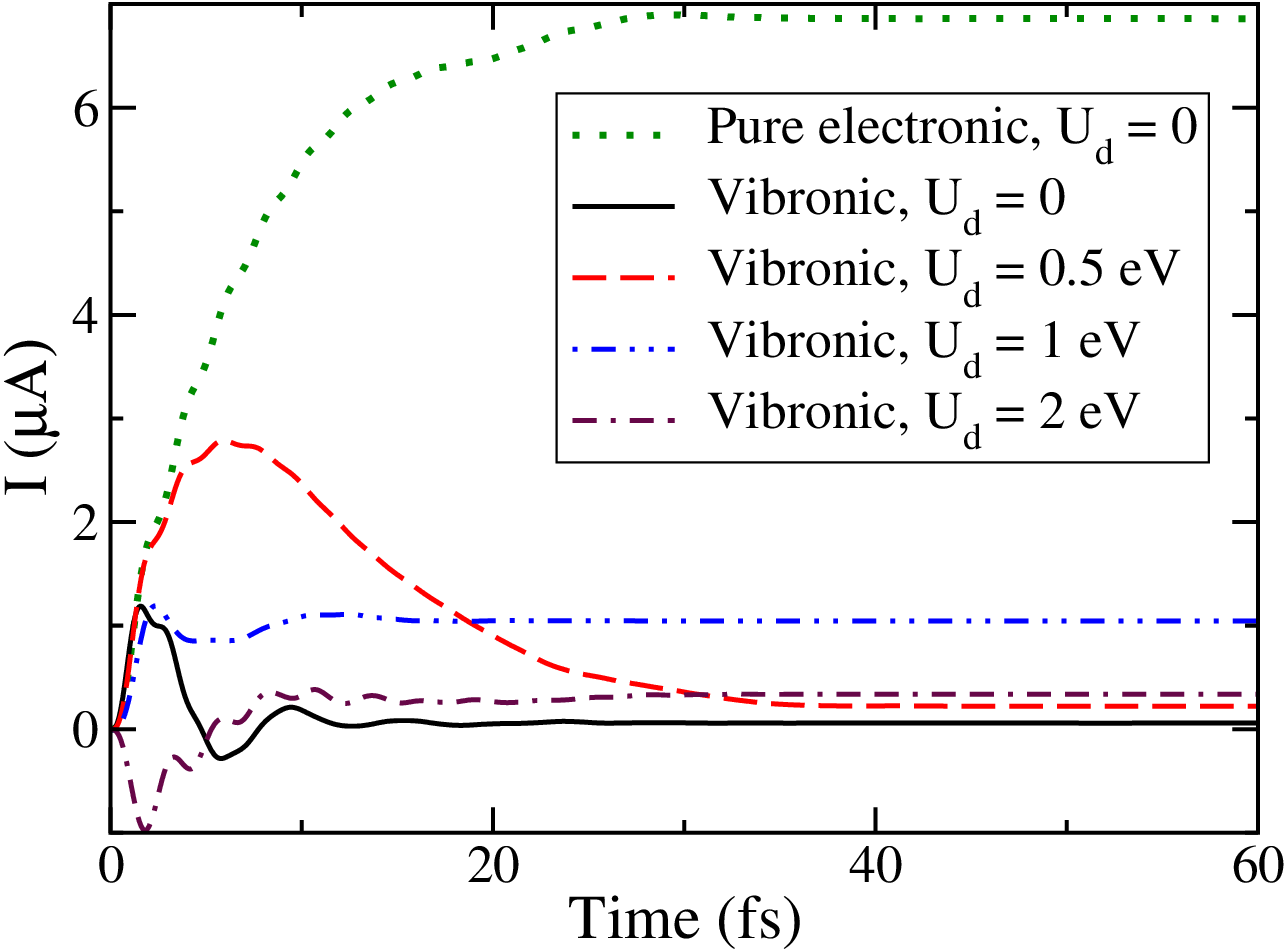}

\begin{flushleft}
(b)
\end{flushleft}
\includegraphics[clip,width=0.45\textwidth]{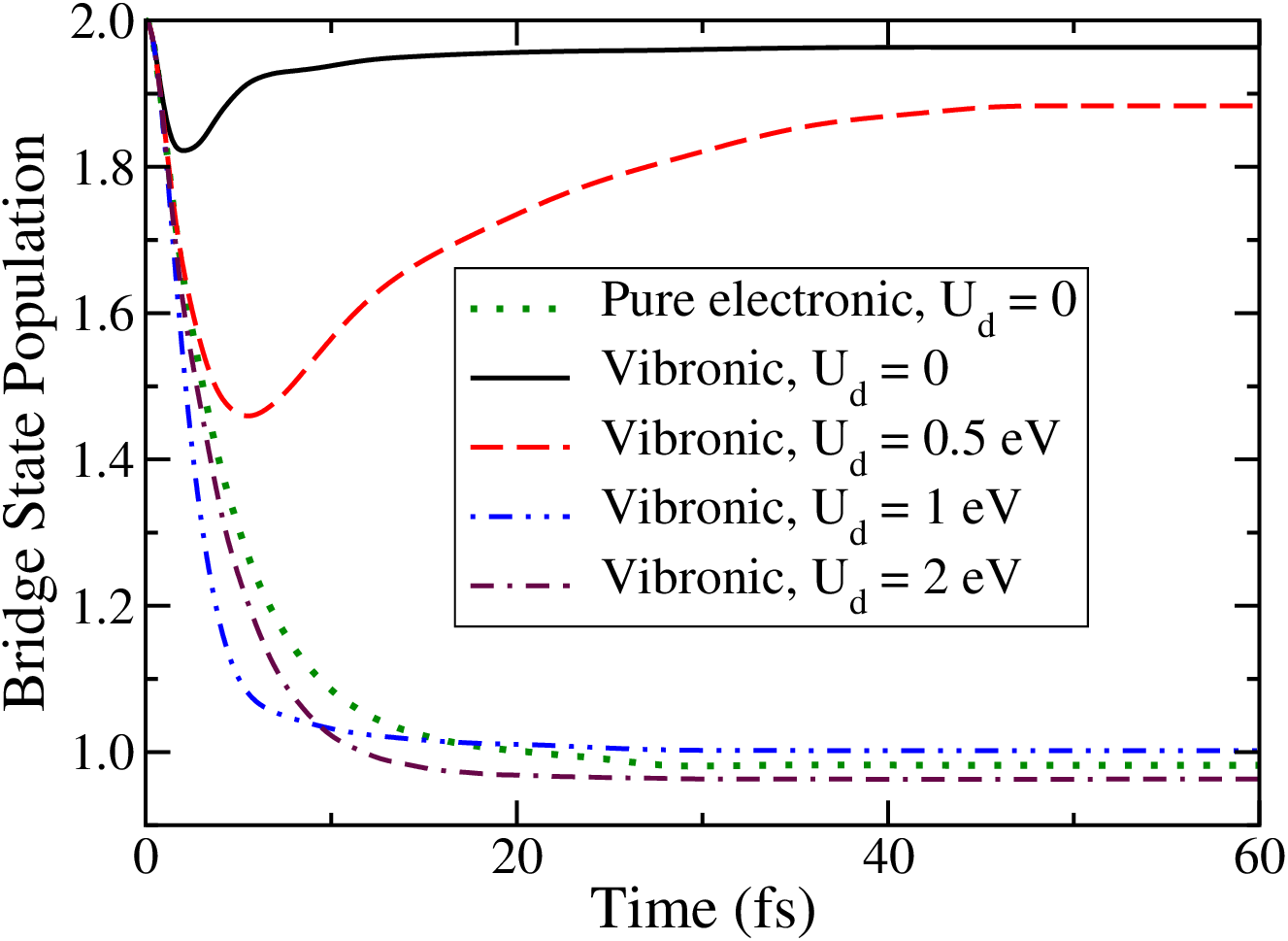}

\caption{(a) Time-dependent current $I(t)$ for different electron-electron
  coupling strength $U_d$  and (b)
the corresponding electronic population of the bridge state.
The results are obtained for a model, which includes both electron-electron  and
electron-vibrational coupling. For comparison, result for a purely electronic
model (i.e.\ $U_d=0$, $\lambda=0$) are shown as indicated in the legend.
The source-drain voltage is $V = 0.1$V. The electronic 
parameters are are: $\alpha_e = 0.2$eV, $\beta_e = 1$eV, $E_d - E_f = 0$. 
The reorganization energy and characteristic frequency for the vibrational bath are $\lambda = 0.25$ eV and 
$\omega_c = 500{\rm cm}^{-1}$, respectively. }
\label{fig7}
\end{figure}

Fig.~\ref{fig7} shows results for a model, where the electronic
parameters are the same as in Fig.~\ref{fig2}, i.e., $\alpha_e = 0.2$ eV, $\beta_e = 1$ eV, 
$E_d - E_f= 0$, and a  source-drain voltage  of $V=0.1$ V. The electronic
degrees of freedom are coupled to a vibrational bath
modeled by an Ohmic spectral density, as described Sec.\ \ref{modeltight}.  The characteristic frequency 
and the reorganization energy of the vibrational bath are $\omega_c = 500$~cm$^{-1}$ and 
$\lambda = 2\alpha\omega_c = 0.25$eV, respectively. These values are typical
for larger molecular systems. 
Without Coulomb repulsion and coupling to
the vibrational bath, this model corresponds to the resonant transport regime. Including the couplings to 
the vibrational modes has a significant impact on the electrical current. After a short transient time 
the coupling to the vibrations becomes effective and results in a suppression of the current. As illustrated
by the solid black line in Fig.~\ref{fig7}, the effect is very pronounced and the stationary current is 
essentially blocked.  The underlying  mechanism can be qualitatively rationalized by considering the
energy level of the bridge states. For any finite bias voltage, the bare energy of the
bridge states ($E_d - E_f = 0$) is located between  the chemical potential of the leads and thus, within 
a purely electronic model, current can flow. The coupling to the vibrations results in a polaron shift 
of the energy of the bridge state given by the reorganization energy
$\lambda$. For electronic-vibrational coupling strengths with  $\lambda > |V|/2$ the
polaron-shifted energy of the bridge state is below the chemical potentials of both leads and thus
current is blocked. This effect, referred to as phonon blockade of the current, has been observed e.g. in 
quantum dots\cite{Weig04} and has been analyzed previously.\cite{Wang11}  
As shown in Fig.~\ref{fig7}(b), the bridge states are almost fully occupied in
this case.

\begin{figure}[!ht]

\includegraphics[clip,width=0.45\textwidth]{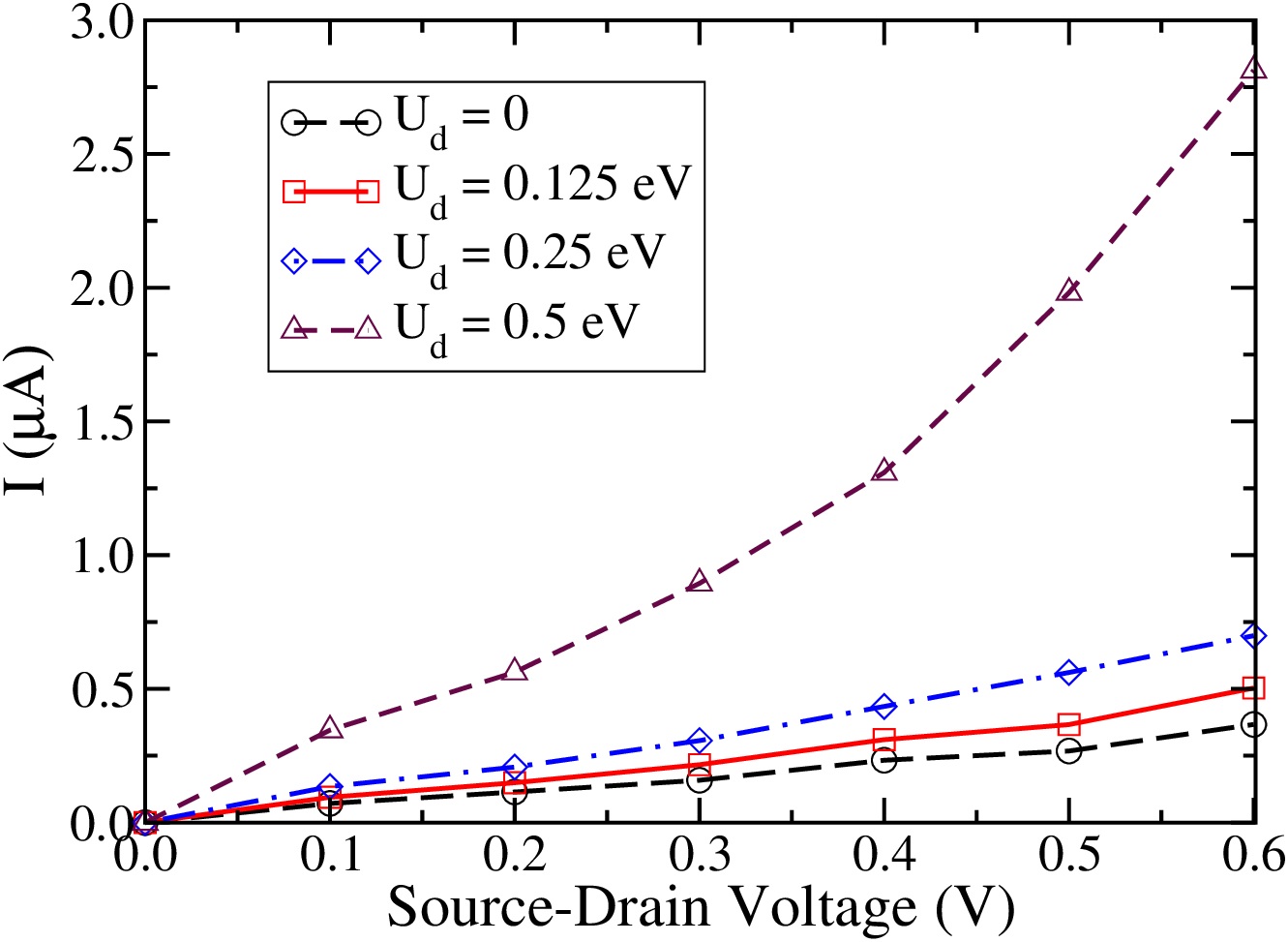}

\caption{Current-voltage characteristics for different electron-electron
  coupling strength $U_d$. The results are obtained for a
  model, which includes both electron-electron and
electron-vibrational coupling. Other parameters are: $\alpha_e = 0.2$eV, $\beta_e = 1$eV, $E_d - E_f = 0$.
The reorganization energy and characteristic frequency for the vibrational bath are $\lambda = 0.25$ eV and 
$\omega_c = 500{\rm cm}^{-1}$, respectively.
The lines are intended as a guide to the eye.}
\label{fig8}
\end{figure}

When the Coulomb repulsion term is included in the simulation (in addition to the  vibrational
bath), the energy level of the doubly occupied bridge state is shifted to
higher energies as discussed for the previous models.  For smaller 
values of $U_d$, this brings the polaron-shifted bridge state back to the conduction window and thus 
increases the stationary current. This can be seen from the currents for $U_d=0.5$eV and $U_d=1$eV in 
Fig.~\ref{fig7}(a) and in Fig.~\ref{fig8}, which shows the current-voltage characteristics for a
few selected values of $U_d$.  It is evident that for small $U_d$ the stationary current increases
versus $U_d$.

However, if $U_d$ becomes too large, e.g., $U_d=2$eV in Fig.~\ref{fig7}(a), the
doubly occupied bridge state has too high energy and is located above the
conduction window, 
which again results in a suppression of the 
stationary current. On the other hand, the overall population of the bridge state decreases monotonically 
upon increase of $U_d$ and reaches a value of unity for large
$U_d$ because then the bridge state can only be singly occupied.
Due to the strongly correlated dynamics in this parameter regime, including both electron-electron and
electronic-vibrational coupling, convergence for larger values of $U_d$ and
larger voltages than those depicted in Figs,~\ref{fig7}, ~\ref{fig8} is
difficult within our present implementation of the ML-MCTDH-SQR methodology.
Experience shows that convergence in this regime can be facilitated by
transforming the current Hamiltonian to another form in order
to reduce correlation effects.  This will be the subject of future work.

Although the interpretation of the above vibronic and electronic correlated transport properties
is appealing in terms of the energetics of the bridge states, it should be emphasized that the mechanism 
involves the formation of correlated  many-body states that are significantly more complex than this  
noninteracting electronic picture, and cannot be fully described by just considering the static shift 
of the energy of the bridge states. This is evident by examining the strength
of the interaction parameters $\lambda$ and $U_d$ 
in Fig.~\ref{fig7}.  Thus, an accurate description of the vibrational and electronic dynamics as well 
as their couplings is essential to obtain a quantitative description of the many-body quantum dynamics and 
the transport characteristics.

\section{Concluding Remarks}\label{conclusions}

In this paper we have employed the ML-MCTDH-SQR method to study correlated electron transport 
through model single-molecule junctions. Extending our previous
work,\cite{wan09:024114,Wang11} we have considered
models which include both electron-electron and electron-vibrational
interaction. The ML-MCTDH-SQR method allows an accurate, in principle 
numerically exact treatment of this many-body quantum transport problem
including both the transient dynamics and the steady state.

The results obtained for selected model systems demonstrate the complex
interplay  of electronic and vibrational
dynamics. For example, strong electron-vibrational coupling may result in a
pronounced suppression of the electrical current (phonon blockade), which is accompanied
by the formation of a polaron-like state. Including electron-electron
interaction,  this suppression of the current can be partially lifted because the transport
channel provided by the doubly occupied bridge state shifts into the
conductance window.

In the present work we have considered a model with a single electronic state at the
molecular bridge. It should be noted, however, that the ML-MCTDH-SQR method can
also be applied to more complex models with various electronic states and
interacting electrons in the leads. In addition to transport problems it may also be used to
describe photoinduced dynamics in molecular adsorbates at metal or semiconductor
surfaces including a proper description of correlation effects.
Another important phenomenon in correlated electron transport is the Kondo
effect.\cite{Hewson93,Wiel00,lia02:725} The application of the methodology to simulate transport in the
Kondo regime, in particular for very small voltage, requires special
discretization techniques (e.g., the scheme pioneered by Wilson\cite{Wilson75}) and can
be facilitated by the use of correlated initial states.  This will be considered in  future work.

\section*{Acknowledgments}
This work has been supported by the National Science Foundation
CHE-1012479 (HW), the 
German-Israeli Foundation for Scientific Development (GIF) (MT), and the 
Deutsche Forschungsgemeinschaft (DFG) through SFB 953 and the Cluster of Excellence ’Engineering of Advanced
Materials’ (MT), and used resources
of the National Energy Research Scientific Computing Center, which is supported by the
Office of Science of the U.S.  Department of Energy under Contract
No. DE-AC02-05CH11231.
MT gratefully acknowledges the hospitality of the IAS at the Hebrew University 
Jerusalem within the workshop on molecular electronics.

\pagebreak

%
%


\clearpage

\clearpage

\clearpage

\clearpage

\clearpage

\vspace{3cm}

\clearpage
~
\vspace{3cm}

\clearpage

\clearpage
~
\vspace{3cm}

\end{document}